\documentclass[aps,pra,showpacs,groupedaddress,floatfix,nofootinbib]{revtex4}
\usepackage{dcolumn}

\begin{document}

\title{Inversion of perturbation series}
\author{Paolo Amore}
\email{paolo@ucol.mx}
\affiliation{Facultad de Ciencias, Universidad de Colima,\\
Bernal D\'{i}az del Castillo 340, Colima, Colima,\\
Mexico.}

\author{Francisco M. Fern\'andez}\email{fernande@quimica.unlp.edu.ar}
\affiliation{INIFTA (Conicet,UNLP), Divisi\'on Qu\'imica Te\'orica, Diag. 113
y 64 S/N, \\
Sucursal 4, Casilla de Correo 16, 1900 La Plata, Argentina}

\begin{abstract}
We investigate the inversion of perturbation series and its resummation, and
prove that it is related to a recently developed parametric perturbation
theory. Results for some illustrative examples show that in some cases
series reversion may improve the accuracy of the results.
\end{abstract}


\maketitle

\section{Introduction\label{sec:intro}}

Perturbation theory yields the solution of a problem in the form of a power
series of a properly chosen model (or dummy) parameter. When the convergence
radius of this series is too small for the physical application, or the
series converges too slowly, it is customary to resort to a resummation
method that produces an approximant that improves the results. There are
many approaches for that purpose; among them we mention Pad\'{e}
approximants, Borel summation, algebraic approximants, continued fractions,
nonlinear transformations of the expansion parameter, etc.~\cite
{BO78,AFC90,F00}.

The purpose of this paper is to investigate the application of a well--known
mathematical method, named series reversion or series inversion~\cite
{A53,MF53,AS72}, to perturbation theory. In Sec.~\ref{sec:inv_ser} we
outline the main ideas of the approach, in Sec.~\ref{sec:param_PT} we show
the connection between the method of series reversion and a recently
developed parametric perturbation theory~\cite{A07a,A07b,A07c}. In Sec.~\ref
{sec:examples} we compare the accuracy of resummation of the direct and
inverse series for some illustrative examples. Finally, in Sec.~\ref
{sec:conclusions} we draw some conclusions about the usefulness of the
approach.

\section{Inversion of Series\label{sec:inv_ser}}

Suppose that we are trying to estimate accurate values of an unknown
function $E(g)$ from a few available coefficients of its power series
\begin{equation}
E(g)=\sum_{j=0}^{\infty }E_{j}g^{j}  \label{eq:g_series}
\end{equation}
When the convergence radius of this series is too small for the physical
application, or the series converges too slowly, it is custommary to resort
to a resummation method that produces an approximant $E\approx
A_{E}^{[N]}(g) $ from the partial sum of order $N$: $S_{N}=E_{0}+E_{1}g+%
\ldots +E_{N}g^{N}$~\cite{BO78,AFC90,F00}.

It is always possible to invert the series (\ref{eq:g_series}) and obtain $g$
in terms of $\Delta E=E-E_{0}$:~\cite{A53,MF53,AS72}
\begin{equation}
g=\Delta E\sum_{j=0}^{\infty }G_{j}\Delta E^{j}=\frac{\Delta E}{E_{1}}-\frac{%
E_{2}\Delta E^{2}}{E_{1}^{3}}+\frac{\left( 2E_{2}^{2}-E_{1}E_{3}\right)
\Delta E^{3}}{E_{1}^{5}}+\ldots  \label{eq:DE_series}
\end{equation}
If we apply a resummation method to this series we obtain an approximation
of the form $g\approx A_{g}^{[N]}(\Delta E)$. This strategy may give more
accurate results than the former one if the radius of convergence or the
region of utility of the series (\ref{eq:DE_series}) is greater than that of
the direct expansion (\ref{eq:g_series}). However, even in this favorable
situation we are paying the price of having the inverse of the desired
function.

In order to get some motivation for the use of the inverse series, consider
the function
\begin{equation}
E=\sqrt{1+g}=1+\frac{1}{2}g-\frac{1}{8}g^{2}+\ldots  \label{eq:trivial_ex}
\end{equation}
that is real for all $g>-1$. The Taylor series about $g=0 $ converges only
for $|g|<1$ because of the branch point at $g=-1$. However, the inverse
series $g=2\Delta E+\Delta E^{2}$ converges for all $\Delta E$ suggesting
that in some cases it is convenient to use the latter instead of the former.

\section{Parametric perturbation theory\label{sec:param_PT}}

In what follows we show that the inversion of series is related to a
recently developed parametric perturbation theory~\cite{A07a,A07b,A07c}. If we
define the parameter $\rho =\Delta E/E_{1}$ then
\begin{equation}
g=\rho \sum_{j=0}^{\infty }G_{j}E_{1}^{j+1}\rho ^{j}=\rho -\frac{E_{2}\rho
^{2}}{E_{1}}+\frac{\left( 2E_{2}^{2}-E_{1}E_{3}\right) \rho ^{3}}{E_{1}^{2}}%
+\ldots  \label{eq:rho_series}
\end{equation}
Application of a resummation method gives an expression for $g$ in terms of
the parameter $\rho $:
\begin{equation}
g\approx A_{g}^{[N]}(E_{1}\rho )  \label{eq:g(rho)}
\end{equation}
which toghether with
\begin{equation}
E = E_{0}+E_{1}\rho  \label{eq:E(rho)}
\end{equation}
define an approximate parametric representation of $E(g)$. These equations
are the basis of the so called parametric perturbation theory developed
recently by Amore~\cite{A07a,A07b,A07c} from a principle of absolute simplicity.

More precisely, in parametric perturbation theory one introduces an
approximant $g=A^{[N]}(\rho )$ into the direct expansion (\ref{eq:g_series})
and choose the approximant coefficients so that $E=b_{0}+b_{1}\rho +[\rho
^{N+1}]$. Since the approximant is constructed so that $A^{[N]}(\rho
)\approx \rho $ for $\rho \ll 1$~\cite{A07a,A07b,A07c} then we realize that $%
b_{0}=E_{0}$ and $b_{1}=E_{1}$, and conclude that parametric perturbation
theory is basically the inversion of the perturbation series.

In order to see the connection clearer consider the trivial example (\ref
{eq:trivial_ex}) again. If we substitute $\rho +c_{2}\rho ^{2}+\ldots
+c_{N}\rho ^{N}$ for $g$ into the series (\ref{eq:trivial_ex}) and choose
the coefficients $c_{j}$ in order to remove the terms of order greater that
one, we obtain $c_{2}=1/4$ and $c_{j}=0$, $j>2$. The resulting expressions $%
E=1+\rho /2$ and $g=\rho +\rho ^{2}/4$ exactly agree with the inverse series
shown above.

In some cases one easily estimates a range of utility of the parametric
representation of $E(g)$. For example, suppose that we know that $dE/dg<0$ ($%
>0$), then $E_{1}<0$ ($>0$). In both cases the range of utility of the
approximant $g\approx A^{[N]}(\rho )$ is determined by the conditions $\rho
>0$ and $dg/d\rho >0$.

\section{Examples\label{sec:examples}}

As an illustrative example we consider the integral
\begin{equation}
E(g)=\int_{0}^{\infty }e^{-x^{2}-gx^{4}}dx.  \label{eq:int_gaus}
\end{equation}
The formal expansion
\begin{equation}
E(g)=\frac{\sqrt{\pi }}{2}\left( 1-\frac{3g}{4}+\frac{105g^{2}}{32}-\frac{%
3465g^{3}}{128}+\frac{675675g^{4}}{2048}-\frac{43648605g^{5}}{8192}+\ldots
\right)  \label{eq:E(g)_gaus_series}
\end{equation}
is known to diverge for all values of $g$~\cite{AFC90} which is reflected by
the form of the expansion coefficients $E_{n}=(-1)^{n}\Gamma
(2n+1/2)/[2(n!)] $. We can rewrite the inverse expansion
\begin{equation}
g(\Delta E)=-\frac{8\Delta E}{3\sqrt{\pi }}+\frac{280\Delta E^{2}}{9\pi }-...
\label{eq:g(DE)_gaus}
\end{equation}
in terms of the parameter $\rho =-3\sqrt{\pi }\Delta E/8$ as
\begin{equation}
g(\rho )=\rho +\frac{35\rho ^{2}}{8}+\frac{35\rho ^{3}}{16}+\frac{17675\rho
^{4}}{256}-\frac{263095\rho ^{5}}{256}+\ldots  \label{eq:g(rho)_series_gaus}
\end{equation}
In this case we have
\begin{equation}
E(\rho )=\frac{\sqrt{\pi }}{2}-\frac{3\sqrt{\pi }\rho }{8}
\label{eq:E(rho)_gaus}
\end{equation}
The pair of equations for $g(\rho )$ and $E(\rho )$ are a parametric
representation of the function $E(g)$.

For simplicity we restrict to series of order five with the purpose of
illustrating some ways of constructing the approximants. In order to
determine the range of applicability of the $\rho $--series we take into
consideration that $dg/d\rho >0$ for all $\rho >0$. Therefore, we assume
that it may be resonable to use the parametric representation for all $%
0<\rho <\rho _{m}$, where $\rho _{m}$ is the smallest positive root of $%
dg/d\rho =0$. In the case of the partial sum of order five (\ref
{eq:g(rho)_series_gaus}) we have $\rho _{m}=0.1659$ that leads to $%
g_{m}=g(\rho _{m})=0.2194$. Table~\ref{tab:series} shows that the parametric
representation embodied in equations (\ref{eq:g(rho)_series_gaus}) and (\ref
{eq:E(rho)_gaus}) yields better results than the direct $g$--power series (%
\ref{eq:E(g)_gaus_series}), at least for all $g<g_{m}$.

We can improve the accuracy of the results by means of resummation methods.
For simplicity we try Pad\'{e} approximants~\cite{BO78,AFC90} built from the
$g$--series and $\rho $--series of order five shown above. Since $%
E(g\rightarrow \infty )=0$ we choose the $[2/3]$ Pad\'{e} approximant
\begin{equation}
E\approx \frac{2\sqrt{\pi }(115460139g^{2}+28532448g+1163200)}{%
141105195g^{3}+534788100g^{2}+117619392g+4652800}  \label{eq:E(g)_gaus_pade}
\end{equation}
The $[2/3]$ Pad\'{e} approximant for the inverse series
\begin{equation}
g=\frac{16\rho (219707\rho +15640)}{26152805\rho ^{3}-11137140\rho
^{2}+2420512\rho +250240}  \label{eq:g(rho)_gaus_pade}
\end{equation}
presents a maximum at $\rho _{m}=0.360$ where $g_{m}=0.607$. Table~\ref
{tab:Pade} shows that the Pad\'{e} approximant for the direct series is more
accurate than the one for the inverse series.

The $[3/2]$ Pad\'{e} approximant
\begin{equation}
g=\frac{\rho \left( 15904+318204\rho +747223\rho ^{2}\right) }{%
15904+248624\rho -375297\rho ^{2}}  \label{eq:g(rho)_gaus_pade_2}
\end{equation}
exhibits a pole at $\rho _{0}=0.721$ and we can therefore use this
parametric representation for all $0<g<\infty $ when $0<\rho <\rho _{0}$.
The main limitation of this approach is that $E(g\rightarrow \infty
)=E_{0}+E_{1}\rho _{0}=0.407$. Table~\ref{tab:Pade_2} shows that the inverse
series is slightly more accurate for small $g$ but leads to a wrong limit.
On the other hand, the direct series approaches the correct limit too fast
as $1/g$.

The approximants so far used do not take into consideration the asymptotic
behavior $E(g)\sim g^{-1/4}$ as $g\rightarrow \infty $. In the case of the
direct series we may calculate $[N/N+1](g)$ Pad\'{e} approximants for $%
E(g)^{4}$ and then use $[N/N+1](g)^{1/4}$ as a reasonable approximation; for
example
\begin{equation}
E\approx \frac{\sqrt{\pi }}{2}\left( \frac{116949g^{2}+27216g+1060}{%
218277g^{3}+190647g^{2}+30396g+1060}\right) ^{1/4}.
\label{eq:E(g)_gaus_pade_asin}
\end{equation}
We do not know how to obtain reasonable approximants for the inverse series;
for that reason we try Amore's approach $g(\rho )=\rho [N/N+1](\rho)^{5}$;
for example,
\begin{equation}
g=\frac{-1048576\rho (24078737127\rho ^{2}+4884321032\rho +161747680)^{5}}{%
(240656732281\rho ^{3}-321691293064\rho ^{2}-75884668992\rho -2587962880)^{5}%
}  \label{eq:g(rho)_gaus_pade_asin}
\end{equation}
constructed from the $\rho $--series of order six. One realizes that this is
the parametric perturbation approach proposed by Amore~\cite{A07a,A07b,A07c} for
this particular case. In the neighbourhood of the pole $\rho=\rho_0$ closest
to origin this approximant behaves as $g\approx K_{0}(\rho -\rho _{0})^{-5}$
so that $E(g)\approx E_{0}+E_{1}\rho _{0}+E_{1}K_{0}^{1/5}g^{-1/5}$. We
appreciate that this parametric representation tends to a wrong limit
(because $E_{0}+E_{1}\rho _{0}\neq 0$ in general) and in a wrong way. Table
\ref{tab:Pade_asin} shows that these approximans are more accurate than the
previous ones, as expected, and that the direct series is clearly better
than the inverse one. Notice that the parametric representation gives the
wrong limit $E(g\rightarrow \infty )=0.861$. Besides, remember that we need
perturbation coefficients of order 5 and 6 of the direct and reverse series,
respectively, in the construction of such approximants.

Following the same philosophy we have also tried $[3/2]$ approximants for
the reverse series

\begin{equation}
g=\frac{\rho \left( 583627586759\rho ^{3}+767002386296\rho
^{2}+44364247040\rho -265333760\right) ^{5}}{32768\left( 290961289397\rho
^{2}+5574551760\rho -33166720\right) ^{5}}
\label{eq:g(rho)_gaus_pade_asin_2}
\end{equation}
Table~\ref{tab:Pade_asin_2} shows that this parametric representation yields
better results than the preceding one, but it also leads to a wrong limit: $%
E(g\rightarrow \infty )=0.883$.

The polylogarithm function
\begin{equation}
Li_{s}(z)=\frac{z}{\Gamma (s)}\int_{0}^{\infty }\frac{t^{s-1}}{e^{t}-z}\,dt
\label{eq:polylog}
\end{equation}
appears in several fields of theoretical physics, for example, in the
Bose--Einstein and Fermi--Dirac distributions~\cite{MQ76}. The Taylor
expansion of this function about $z=0$ yields the series
\begin{equation}
Li_{s}(z)=\sum_{n=1}^{\infty }\frac{z^{n}}{n^{s}}  \label{eq:polylog_series}
\end{equation}
In this case the inverse series gives more accurate results than the direct
one and, consequently, we draw the same conclusion regarding the Pad\'{e}
approximants constructed from them. Table~\ref{tab:polylog} shows that the
Pad\'{e} approximant $[5/6](\rho )$ for the former series gives considerably
more accurate results than the $[6/7](g)$ Pad\'{e} approximant for the
latter one.

The formal Taylor series about $g=0$ for the function
\begin{equation}
E(g)=\int_{0}^{\infty }\frac{e^{-x}}{1+gx}\,dx  \label{eq:E(g)_exp_int}
\end{equation}
is divergent as shown by its coefficients $E_{n}=(-1)^{n}n!$. The asymptotic
behavior at $g\gg 1$ is given by
\begin{equation}
E(g)\approx \frac{\ln (g)-\gamma }{g}+\frac{\ln (g)-\gamma +1}{g^{2}}+\ldots
\label{eq:exp_int_large_g}
\end{equation}
where $\gamma =0.5772156649\ldots $ is Euler's constant. In this case the
inverse series is more accurate than the direct one, and exactly the same
situation takes place for their corresponding Pad\'{e} approximants.

As another illustrative example consider the function $E(g)$ defined by $%
g=Ee^{-E}$. The power series $E(g)=g+g^{2}+3g^{2}/2+\ldots $ converges for $%
g<e^{-1}$ which is reflected by the form of the expansion coefficients $%
E_{n}=n^{n-1}/n!$, $n\geq 1$. On the other hand, the inverse series $%
g=E-E^{2}+\ldots +(-1)^{n-1}E^{n}/(n-1)!+\ldots $ converges for all $E$ and
is therefore preferable for practical applications. However, it is not
always true that the inverse series have greater convergence radius than the
direct one; simply consider the function $E(g)=ge^{-g}$, where the role of
the variables has been reversed. In this case parametric perturbation theory
will perform poorer than standard perturbation theory.

\section{Conclusions\label{sec:conclusions}}

In this paper we investigated the usefulness of the inversion of
perturbation series and its resummation. Our simple examples show that in
some cases it is convenient to resort to the inverse series, but in others
the straightforward perturbation series is expected to provide better
results. We have also proved that the recently proposed parametric
perturbation theory~\cite{A07a,A07b,A07c}, based on the principle of absolute simplicity, consists
of a convenient modification and resummation of the inverse series.
Therefore, in some cases this approach will not perform better than
well--known resummation methods on the direct series. In addition to it,
parametric perturbation theory exhibits two disadvantages: producing the
inverse of the desired function, and the difficulty of deriving the correct
asymptotic limit and behaviour when it is known.

\begin{table}[H]
\caption{Series expansions of order five for the integral~(\ref{eq:int_gaus}%
) }
\label{tab:series}
\begin{center}
\begin{tabular}{D{.}{.}{2}D{.}{.}{11}D{.}{.}{11}D{.}{.}{11}D{.}{.}{11}}
\hline
\multicolumn{1}{c}{$g$}& \multicolumn{1}{c}{$\rho$}&
\multicolumn{1}{c}{Eq.(\ref{eq:E(g)_gaus_series})} &
\multicolumn{1}{c}{Eq.(\ref{eq:g(rho)_series_gaus})} & \multicolumn{1}{c}{Exact} \\
\hline
0.04& 0.227268254&  0.8630223905&   0.8632172917&  0.8632281022\\
0.08& 0.2211979138& 0.8358839701&   0.8447690498&  0.8449941504\\
0.12& 0.213865426&  0.7500155805&   0.8285524745&  0.8296883134\\
0.16& 0.2043179764& 0.4525374722&   0.7504228564&  0.8164228001\\
0.2&  0.1891655735& -0.3655376234&  0.7604942069&  0.8046805576
\end{tabular}
\par
\end{center}
\end{table}

\begin{table}[H]
\caption{Pad\'e approximants [2/3]for the integral~(\ref{eq:int_gaus})}
\label{tab:Pade}
\begin{center}
\par
\begin{tabular}{D{.}{.}{2}D{.}{.}{11}D{.}{.}{11}D{.}{.}{11}D{.}{.}{11}D{.}{.}{11}}
\hline
\multicolumn{1}{c}{$g$}& \multicolumn{1}{c}{$\rho$}&
\multicolumn{1}{c}{Eq.(\ref{eq:E(g)_gaus_pade})} &
\multicolumn{1}{c}{Eq.(\ref{eq:g(rho)_gaus_pade})} & \multicolumn{1}{c}{Exact} \\
\hline
0.1& 0.07419851329& 0.836884189& 0.8369093852& 0.8370429277\\
0.2& 0.124755604& 0.803286224& 0.8033055939& 0.8046805576\\
0.3& 0.166889096& 0.7760554974& 0.7753007175& 0.7800434542\\
0.4& 0.2069906611& 0.7523625938& 0.7486464025& 0.7600358198\\
0.5& 0.2515059286& 0.7310124346& 0.719058431& 0.7431554088\\
0.6& 0.392980645& 0.7113938594& 0.6250244038& 0.7285439429
\end{tabular}
\par
\end{center}
\end{table}

\begin{table}[H]
\caption{Pad\'e approximant [2/3] for the $g$--series and [3/2] for the
inverse series of the integral~(\ref{eq:int_gaus})}
\label{tab:Pade_2}
\begin{center}
\par
\begin{tabular}{D{.}{.}{2}D{.}{.}{11}D{.}{.}{11}D{.}{.}{11}D{.}{.}{11}D{.}{.}{11}}
\hline
\multicolumn{1}{c}{$g$}& \multicolumn{1}{c}{$\rho$}&
\multicolumn{1}{c}{Eq.(\ref{eq:E(g)_gaus_pade})} &
\multicolumn{1}{c}{Eq.(\ref{eq:g(rho)_gaus_pade_2})} & \multicolumn{1}{c}{Exact} \\
\hline
0.1& 0.07404520814& 0.836884189& 0.8370112825& 0.8370429277\\
1& 0.3079042644& 0.6446831837& 0.6815721382& 0.6842134278\\
10& 0.6043806547& 0.2142628036& 0.4845131183& 0.4609804743\\
100& 0.7061454902& 0.02801405374& 0.4168730654& 0.2772884009\\
1000& 0.7196727352& 0.002890400973& 0.4078819088& 0.1594808649\\
10000& 0.7210732433& 0.000289961386& 0.4069510328& 0.09033502245
\end{tabular}
\par
\end{center}
\end{table}

\begin{table}[H]
\caption{Improved Pad\'e approximants [2/3] for the integral~(\ref
{eq:int_gaus})}
\label{tab:Pade_asin}
\begin{center}
\par
\begin{tabular}{D{.}{.}{2}D{.}{.}{11}D{.}{.}{11}D{.}{.}{11}D{.}{.}{11}}
\hline
\multicolumn{1}{c}{$g$}& \multicolumn{1}{c}{$\rho$}&
\multicolumn{1}{c}{Eq.(\ref{eq:E(g)_gaus_pade_asin})} &
\multicolumn{1}{c}{Eq.(\ref{eq:g(rho)_gaus_pade_asin})} & \multicolumn{1}{c}{Exact} \\
\hline
0.1& 0.07356402668& 0.8369445716& 0.8373311095& 0.8370429277\\
1& 0.2892446121& 0.6715809835& 0.6939746529& 0.6842134278\\
10& 0.5828537448& 0.4198251659& 0.4988214137& 0.4609804743\\
100& 0.03723484095& 0.2393860646& 0.8614780364& 0.2772884009\\
1000& 0.03737319161& 0.1348101326& 0.8613860789& 0.1594808649\\
10000& 0.03737319161& 0.07582023369& 0.8613860789& 0.09033502245
\end{tabular}
\par
\end{center}
\end{table}

\begin{table}[H]
\caption{Improved Pad\'e approximant $[3/2]$ for the inverse series of the
integral~(\ref{eq:int_gaus}) }
\label{tab:Pade_asin_2}
\begin{center}
\par
\begin{tabular}{D{.}{.}{2}D{.}{.}{11}D{.}{.}{11}}
\hline
\multicolumn{1}{c}{$g$}& \multicolumn{1}{c}{$\rho$}&
\multicolumn{1}{c}{Eq.(\ref{eq:g(rho)_gaus_pade_asin_2})}  \\
\hline
0.1&   0.07404345742&  0.8370124462 \\
1&     0.3170281954&   0.6755077332       \\
10&    0.8091513837&   0.3484081181      \\
100&   0.00545012921&  0.882604387     \\
1000&  0.00545012921&  0.882604387    \\
10000& 0.00545012921&  0.882604387
\end{tabular}
\end{center}
\end{table}

\begin{table}[H]
\caption{Values of $Li_{3/2}(g)$ for $g$ close to 1 obtained from Pad\'e
approximants on the direct and inverse series. Exact figures are underlined
for comparison.}
\label{tab:polylog}
\begin{center}
\par
\begin{tabular}{D{.}{.}{6}D{.}{.}{2}D{.}{.}{2}D{.}{.}{11}}
\hline
\multicolumn{1}{c}{$g$}  & \multicolumn{1}{c}{$[6/7](g)$}&
\multicolumn{1}{c}{$[5/6](\rho)$} &
 \multicolumn{1}{c}{Exact}   \\
\hline
0.999999& $ \underline{2}.380740506 $ &  $ \underline{2.608}744256$ & 2.608831900     \\
0.99999 & $ \underline{2}.380591082 $ &  $ \underline{2.6011}53011$ & 2.601179942     \\
0.9999  & $ \underline{2}.379099267 $ &  $ \underline{2.5770}63920$ & 2.577071427     \\
0.999   & $ \underline{2}.364418183 $ &  $ \underline{2.50170}6883$ & 2.501708465     \\
0.99    & $ \underline{2.2}37103024 $ &  $ \underline{2.2716}59944$& 2.271660077     \\
0.9     & $ \underline{1.614}336255 $ &  $ \underline{1.61443852}8$& 1.614438529     \\

\end{tabular}
\end{center}
\end{table}

\end{document}